\documentclass[10pt]{iopart}

\input psfig.sty

\usepackage{graphicx}

\begin{document}

\title{Non-singular inflation with vacuum decay}

\author{Saulo Carneiro}

\address{Instituto de F\'{\i}sica, Universidade Federal da Bahia,
Salvador, BA, Brazil \\ International Centre for Theoretical
Physics, Trieste, Italy\footnote{Associate member}}

\ead{saulo@fis.ufba.br}

\begin{abstract}
On the basis of a semi-classical analysis of vacuum energy in an
expanding spacetime, we describe a non-singular cosmological model
in which the vacuum density decays with time, with a concomitant
production of matter. During an infinitely long period we have an
empty, inflationary universe, with $H \approx 1$. This primordial
era ends in a fast phase transition, during which $H$ and
$\Lambda$ decrease to nearly zero in a few Planck times, with
release of a huge amount of radiation. The late-time scenario is
similar to the standard model, with the radiation phase followed
by a long dust era, which tends asymptotically to a de Sitter
universe, with vacuum dominating again. An analysis of the
redshift-distance relation for supernovas Ia leads to cosmological
parameters in agreement with other current estimations.
\end{abstract}


\section{Introduction}

The role of vacuum energy in cosmology has acquired a renewed
importance, with current observations pointing to the existence of
a negative-pressure component in the cosmic fluid. This has
reinforced the cosmological constant problem \cite{Sahni}, whose
origin can be understood if we write the energy density associated
to vacuum quantum fluctuations. In the case of a massless scalar
field, it is given by the divergent integral
\begin{equation} \label{Lambdanua}
\Lambda_0 \approx \int_0^{\infty} \omega^3 d\omega.
\end{equation}

This integral can be regularized by imposing a cutoff $m$ in the
superior limit of integration, leading to $\Lambda_0 \approx m^4$.
The same result can be derived by introducing a bosonic
distribution function in (\ref{Lambdanua}),
\begin{equation}\label{Lambdareg}
\Lambda_0 \approx \int_0^{\infty}
\frac{\omega^3\,d\omega}{e^{\omega/m}-1}\approx m^4,
\end{equation}
which is equivalent to consider the vacuum fluctuations thermally
distributed, at a characteristic energy $m$.

For a cutoff of the order of the Planck mass, this leads to a vacuum
density $120$ orders of magnitude above the currently observed dark
energy density. One may argue that $m$ should in fact be much
smaller, because vacuum fluctuations above the energy scale of the
QCD phase transition - the latest cosmological vacuum transition -
would lead to quark deconfinement. However, even with this value for
$m$, we obtain a vacuum density $40$ orders of magnitude above the
observed value.

The cosmological constant problem may be alleviated by the
following reasoning. The above energy density is obtained in a
flat spacetime, but in such a background the energy-momentum
tensor in Einstein equations should be zero. Therefore, by
consistence, the above vacuum density must be exactly canceled by
a bare cosmological constant. If we now derive the vacuum
contribution in a curved spacetime, we should expect, after the
subtraction of the bare cosmological constant, a renormalized,
curvature-dependent vacuum energy density. For instance, in the
case of an expanding, spatially isotropic and homogeneous
spacetime, filled with vacuum and matter components, the
renormalized $\Lambda$ should be time dependent, being very high
at early times, but decreasing to zero or nearly zero as the
universe expands \cite{Schutzhold}.

To have an idea of what a varying cosmological term may be, let us
initially consider the case of a de Sitter spacetime. It is
generally believed that the de Sitter cosmological horizon has an
associated temperature given by $H/2\pi$, where $H=\dot{a}/a$ is the
Hubble parameter \cite{Bousso}\footnote{This result was derived by
Gibbons and Hawking on the basis of Euclidian methods
\cite{Gibbons}, but the positiveness of the de Sitter temperature
depends on some appropriate physical interpretation (see, for
example, \cite{Spradlin,Pad}).}. Therefore, a phenomenological
expression for the effective vacuum density in this case may be
derived by substituting $m + H$ for the energy scale $m$ in
(\ref{Lambdareg}), with a subsequent subtraction of $\Lambda_0
\approx m^4$. In this way we obtain
\begin{equation}\label{Lambda}
\Lambda \approx (m+H)^4-m^4.
\end{equation}

In the limit $H>>m$, this leads to the cutoff-independent result
$\Lambda \approx H^4$. Since, in a de Sitter spacetime, $\Lambda
\approx H^2$, this implies that $H \approx 1$, that is, this limit
necessarily describes a de Sitter universe with horizon radius of
the order of the Planck length. On the other hand, in the limit
$H<<m$ we have $\Lambda \approx m^3H$. Using again $\Lambda
\approx H^2$, we obtain $H \approx m^3$, or $\Lambda \approx m^6$.
If, as discussed above, we choose $m$ as the energy scale of the
QCD vacuum phase transition (of the order of the pion mass), the
first result is an expression of Dirac's large number coincidence
\cite{Mena}, while the last one gives approximately the current
observed value of $\Lambda$ \cite{GRF}.

The above discussion concerns stationary spacetimes, for which $H$
and $\Lambda$ are truly constants. Nevertheless, it suggests the
possibility of a universe evolving from an initial, asymptotically
de Sitter phase, with $\Lambda \sim 1$, to a final, asymptotically
de Sitter phase with $\Lambda \sim m^6 << 1$, with $\Lambda$
decreasing with time according to (\ref{Lambda}). Before verifying
such a possibility in the following sections, let us briefly discuss
the energy conservation in this context.

There is a common belief about the impossibility of a varying
cosmological term, because of the Bianchi identities and the
covariant conservation of the energy-momentum of matter. Indeed, the
Bianchi identities $G^{\mu}_{\nu;\mu}=0$ ($G$ is the Einstein
tensor) imply, via Einstein's equations, the conservation of the
total energy-momentum tensor, $T^{\mu}_{\nu;\mu} =
\tilde{T}^{\mu}_{\nu;\mu} + g^{\mu}_{\nu} \Lambda_{,\mu} = 0$, where
$\tilde{T}$ is the energy-momentum tensor of matter. If one assumes
the independent conservation of matter, i.e.,
$\tilde{T}^{\mu}_{\nu;\mu}=0$, it follows that $\Lambda_{,\mu} = 0$,
that is, $\Lambda$ is a constant.

However, if matter is not independently conserved, $\Lambda$ may
vary with time \cite{Barrow}. In the realm of a FRW spacetime, the
conservation of the total energy-momentum tensor reduces to the form
\begin{equation} \label{continuidade}
\dot{\rho}_T + 3H(\rho_T+p_T)=0,
\end{equation}
where $\rho_T$ and $p_T$ are the total energy density and
pressure, respectively. By using $\rho_T = \rho_m + \Lambda$ and
$p_T = p_m - \Lambda$ ($\rho_m$ and $p_m$ refer to the
corresponding matter quantities), we have
\begin{equation} \label{continuidade2}
\dot{\rho}_m+3H(\rho_m+p_m)=-\dot{\Lambda}.
\end{equation}
In other words, the vacuum decay is concomitant with a process of
matter production, in order to preserve the covariant conservation
of the total energy.

\section{The early times}

In the spatially flat case, the Friedmann equations give $\rho_T =
3H^2$. In the limit of very early times we can take $H>>m$, and
(\ref{Lambda}) reduces to $\Lambda = 3H^4$, where the factor $3$
is not important and was chosen for mathematical convenience.
Using for matter the equation of state of radiation, $p_m =
\rho_m/3$, the conservation equation (\ref{continuidade2}) then
leads to the evolution equation
\begin{equation}\label{evolucao}
\dot{H}+2H^2-2H^4=0.
\end{equation}

For $0<H<1$, the solution of (\ref{evolucao}) is given by
\begin{equation}\label{H}
2t=\frac{1}{H}-\tanh^{-1}H,
\end{equation}
where $t$ is the cosmological time, and an integration constant
was conveniently chosen.

\begin{figure}
\psfig{figure=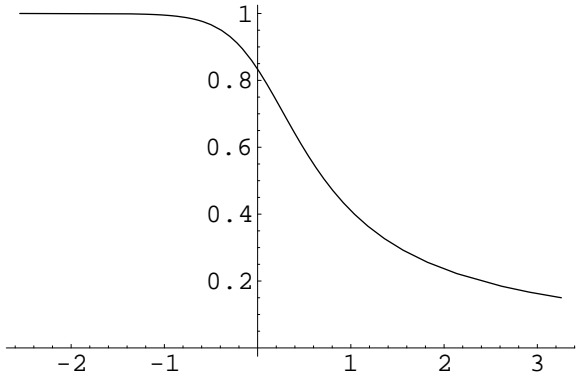} \caption{The Hubble parameter as a
function of time (in Planck units)}
\end{figure}

This solution is plotted in Figure 1, with $t$ and $H$ expressed in
Planck units. We can see that this universe has no initial
singularity, existing since an infinite past, when it approaches
asymptotically a de Sitter state with $H = 1$. During an infinitely
long period we have a quasi-de Sitter universe, with $H\approx
\Lambda \approx 1$. However, at a given time (arbitrarily chosen
around $t = 0$), the expansion undertakes a fast and huge phase
transition, with $H$ and $\Lambda$ decreasing to nearly zero in a
few Planck times\footnote{The time at which the transition occurs
depends on the integration constant in (\ref{H}). Note, however,
that it takes place a definite time before the present time. In this
sense, we can attribute an age for the subsequent universe, which
will be determined below.}.

\begin{figure}
\psfig{figure=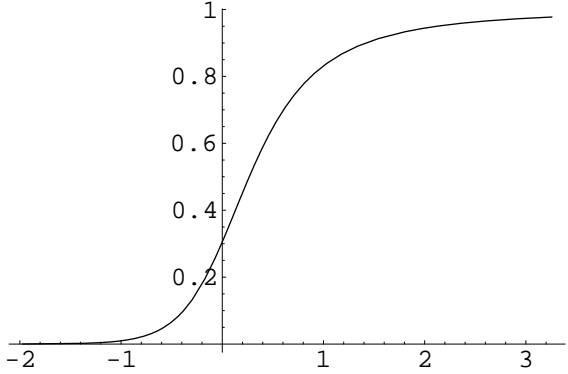} \caption{The relative energy density of
radiation as a function of time}
\end{figure}

From $\rho_m = \rho_T - \Lambda$ we can obtain $\Omega_m = 1 -
H^2$, where $\Omega_m = \rho_m/3H^2$ is the relative density of
matter. Its time variation is plotted in Figure 2. One can see
that $\Omega_m$ changes suddenly during the phase transition, from
nearly zero, in the quasi-de Sitter phase, to nearly $1$ at the
end of the transition. In this way, the present solution has some
attributes of an inflationary universe, with an infinitely long
period of inflation ending in a fast transition during which the
vacuum decays, releasing a huge amount of energy in form of
radiation and relativistic matter. After the transition we have a
radiation-dominated FRW universe, whose subsequent evolution, as
we will see, is similar to the standard $\Lambda$CDM recipe.

\section{Late times}

In the opposite limit we can take $H<<m$, and (\ref{Lambda})
reduces to $\Lambda \approx m^3 H$. Let us write it as $\Lambda =
\sigma H$, with $\sigma \approx m^3$, and let us introduce the
equation of state of matter, $p_m = (\gamma-1) \rho_m$. From
(\ref{continuidade2}) we now have
\begin{equation} \label{evolucao2}
2\dot{H} + 3\gamma H^2 - \sigma \gamma H = 0.
\end{equation}

For $H>0$ and $\rho_m>0$, one has the solution \cite{Borges}
\begin{equation} \label{a}
a(t) = C \left[\exp\left(\sigma \gamma t/2\right) -
1\right]^{\frac{2}{3\gamma}},
\end{equation}
where $C$ is an integration constant, and a second one was
conveniently chosen.

From it we can derive $H(a)$ and, with the help of $\rho_m = 3H^2
- \Lambda$, we obtain
\begin{equation}\label{rhoa}
\rho_m =
\frac{\sigma^2}{3}\left(\frac{C}{a}\right)^{3\gamma/2}\left[1 +
\left(\frac{C}{a}\right)^{3\gamma/2}\right],
\end{equation}
\begin{equation}\label{Lambdaa}
\Lambda = \frac{\sigma^2}{3}\left[1 +
\left(\frac{C}{a}\right)^{3\gamma/2}\right].
\end{equation}

\subsection{The radiation era}

In the radiation phase, doing $\gamma = 4/3$ and taking the limit
$\sigma t<<1$, we have
\begin{equation}\label{asmall}
a \approx \sqrt{2C^2\sigma t/3},
\end{equation}
\begin{equation}\label{rhosmall}
\rho_m = \frac{\sigma^2 C^4}{3a^4} = \frac{3}{4t^2},
\end{equation}
and
\begin{equation}\label{Lambdasmall}
\Lambda = \frac{\sigma^2 C^2}{3a^2} = \frac{\sigma}{2t}.
\end{equation}

The two first results are the same we obtain in the standard
model. The third one shows that, in this limit, $\Lambda$ is
sub-dominant compared to $\rho_m$, and, therefore, the matter
production can be neglected. This is the reason why radiation is
conserved, with an energy density scaling with $a^{-4}$.

Equations (\ref{asmall}) and (\ref{rhosmall}) guarantee that
physical processes taking place during the radiation phase are not
affected by the vacuum decay. For example, the primordial
nucleosynthesis remains unchanged, since the expansion and
reaction rates are the same as in the standard context.

\subsection{The matter era}

In the case of a matter fluid dominated by dust, taking $\gamma =
1$ we obtain, from (\ref{a}),
\begin{equation} \label{adust}
a(t) = C \left[\exp\left(\sigma t/2\right) - 1\right]^{2/3}.
\end{equation}

In the limit $\sigma t<<1$, it reduces to
\begin{equation}\label{adustsmall}
a(t) = C(\sigma t/2)^{2/3},
\end{equation}
that is, the scale factor evolves as in a dust-dominated
Einstein-de Sitter universe.

On the other hand, for $t \rightarrow \infty$ we have
\begin{eqnarray} \label{alarge}
a(t) &=& C \exp\left(\sigma t/3\right),
\end{eqnarray}
i.e., our solution tends asymptotically to a de Sitter universe,
with $H=\sigma/3$.

For the matter and vacuum energy densities, we obtain, from
(\ref{rhoa}) and (\ref{Lambdaa}),
\begin{equation}\label{rhodust}
\rho_m = \frac{\sigma^2 C^3}{3a^3} + \frac{\sigma^2
C^{3/2}}{3a^{3/2}},
\end{equation}
\begin{equation}\label{Lambdadust}
\Lambda = \frac{\sigma^2}{3} + \frac{\sigma^2 C^{3/2}}{3a^{3/2}}.
\end{equation}

It is not difficult to interpret these results. The first terms in
these equations are the standard expressions for the scaling of
matter and the cosmological term, valid if there is no matter
production. The first term of (\ref{rhodust}) is dominant in the
limit $a/C<<1$, when $a$ scales as in (\ref{adustsmall}). The
first term of (\ref{Lambdadust}), on the other hand, is dominant
for $a \rightarrow \infty$, acting as a genuine cosmological
constant.

The second terms in (\ref{rhodust}) and (\ref{Lambdadust}) are
related to the vacuum decay. For large times the matter density
decreases slower then in the $\Lambda$CDM model, because of the
matter production.

\subsection{Supernova constraints}

\begin{figure}
\psfig{figure=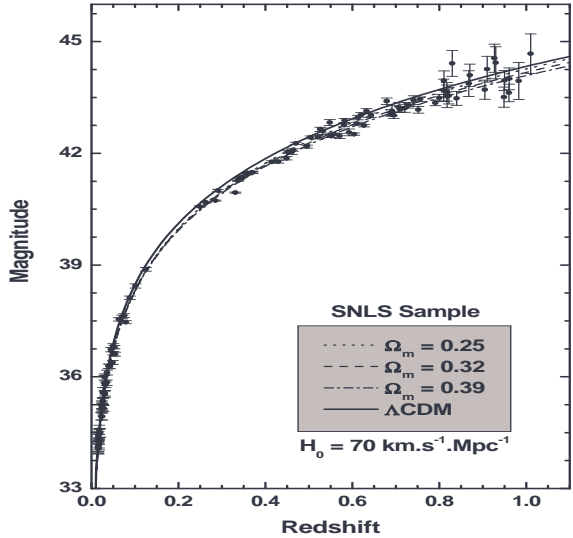,width=3.2truein,height=3.1truein}
\caption{Hubble diagram for 115 supernovae from SNLS Collaboration
\cite{Alcaniz}. The curves correspond to $H_0 = 70\,
\rm{Km.s^{-1}.Mpc^{-1}}$, and selected values of $\Omega_{\rm{m}}$.
For the sake of comparison, the flat $\Lambda$CDM scenario with
$\Omega_{\rm{m}} = 0.27$ is also shown.}
\end{figure}

As we have seen, the cosmological scenario presented here is
similar, on a qualitative level, to the standard scenario of
cosmic evolution. Nevertheless, the vacuum decay constitutes a
substantial difference at late times, and we should verify its
consequences for the quantitative determination of cosmological
parameters like the relative density of matter and the universe
age, for example. For this purpose, the analysis of the
redshift-distance relation for supernovas Ia is of particular
importance \cite{Alcaniz}.

From (\ref{adust}), it is easy to derive the Hubble parameter as a
function of the redshift $z = a_0/a - 1$, where $a_0$ is the
present value of the scale factor. One obtains
\begin{equation} \label{Hz}
H(z) = H_0 \left[1 - \Omega_{\rm{m}} +\Omega_{\rm{m}} (1 +
z)^{3/2}\right],
\end{equation}
where $H_0$ and $\Omega_m$ are the present values of the Hubble
parameter and of the relative density of matter, respectively.

One can see that, as in the $\Lambda$CDM model, we have two
parameters to be adjusted by fitting the supernova data. In order
to do it, we have used (\ref{Hz}) to fit the data of the Supernova
Legacy Survey (SNLS) Collaboration \cite{SNLS}. The result is
shown in Figure 3, where we have plotted our theoretical
redshift-distance relation for three different values of
$\Omega_m$, together with the theoretical relation predicted by
the spatially flat $\Lambda$CDM model with $\Omega_m = 0.27$. In
all cases we have used $H_0 = 70$ (Km/s)/Mpc.

The best fit obtained with (\ref{Hz}) is given by $\Omega_m =
0.32$ and $H_0 = 70$ (Km/s)/Mpc, with $\chi_r^2 = 1.0$. With
$95\%$ of confidence level, we have $0.27 \le \Omega_{\rm{m}} \le
0.37$ and $0.68 \le h \le 0.72\;$ ($h \equiv H_0/100
\rm{Km.s^{-1}.Mpc^{-1}})$.

With these results we can estimate the universe age in this model.
From (\ref{adust}) we can derive an age parameter given by
\begin{equation} \label{age}
H_0 t_0 = \frac{\frac{2}{3}\ln(\Omega_{\rm{m}})}{\Omega_{\rm{m}} -
1}.
\end{equation}
By using the obtained values of $H_0$ and $\Omega_m$, one has $t_0
\simeq 15.7$ Gyr, corresponding to an age parameter $H_0 t_0 =
1.12$.

It is also possible to obtain from (\ref{adust}) the present
deceleration parameter, whose best value is $q = - 0.52$. The
redshift of transition between the decelerated phase and the
accelerated one is $z = 1.62$ \cite{Alcaniz}, showing that we have
a decelerated phase long enough to permit structure formation.

\section{Conclusions}

We have described, on the basis of a macroscopic approach to
vacuum dynamics, a non-singular cosmological scenario in agreement
with our general standard view about the universe evolution. We
have an initially empty, inflationary spacetime, which is driven
to a radiation-dominated phase through a fast phase transition
during which a huge amount of energy is released at the expenses
of the vacuum decay. The radiation phase is indistinguishable from
the standard one, and it is followed by a matter-dominated era,
which evolves to a final de Sitter phase, with vacuum dominating
again.

This scenario is also consistent with a quantitative analysis of
the observed Hubble diagram for supernovas of high redshifts,
leading to cosmological parameters in accordance with other -
non-cosmological - estimations. In particular, the matter density
and age parameters are in the intervals imposed, respectively, by
dynamical estimations of dark matter \cite{calb} and globular
clusters observations \cite{age}.

The reader may object that our matter density parameter is above
the values estimated on the basis of current observations of the
cosmic background radiation \cite{cbr} and barion acoustic
oscillations \cite{bao} - despite the superposition of the
intervals of allowed values. Nevertheless, it must be emphasized
that such estimations are dependent on the adopted cosmological
model, and should be redone in our case. In particular, the
production of matter modifies the standard relations between the
dynamic parameters at the time of last scattering and the present
matter density. An analysis of such observations in the context of
the present model is in progress.

Another point to be analyzed is the evolution of density
perturbations, which could be modified by the matter production. A
preliminary investigation has shown no important difference in the
growing of the density contrast in the matter era, while in the
radiation phase the matter production is irrelevant, as discussed
above.

\section*{Acknowledgments}

I would like to thank Joan Sol\`a and the other organizers of IRGAC
2006 for the warm hospitality in Barcelona, and two anonymous
referees for constructive remarks. This work is partially supported
by CNPq (Brazil).

\end{document}